\documentclass{webofc}
\usepackage[varg]{txfonts}   
\begin{document}

\title{Optimized simulations of $^{50}$Ti(p,$\alpha$) and $^{49}$Ti(d,$\alpha$) reactions for hospital-cyclotron production of $^{47}$Sc}

\author{
\firstname{F.} \lastname{Barbaro}\inst{1,2}\fnsep\thanks{\email{francesca.barbaro@pd.infn.it}}
\and
\firstname{L.} \lastname{Canton}\inst{1}
\and 
\firstname{Y.} \lastname{Lashko}\inst{1,3}
\and
\firstname{L.} \lastname{Zangrando}\inst{1}
}
\institute{INFN, Sezione di Padova, Padova, Italy
\and
Dipartimento di Fisica dell’Università di Pavia, Pavia, Italy
\and
Bogolyubov Institute for Theoretical Physics, Kyiv, Ukraine}

%
%

\abstract{%
The production of $^{47}$Sc, a promising radioisotope for targeted radionuclide therapy, by means of hospital-cyclotron reactions is investigated. Two possible routes are considered: the proton-induced reaction on enriched $^{50}$Ti targets and the deuteron-induced reaction on enriched $^{49}$Ti targets. The cross sections of the reactions are calculated using the TALYS code with optimized parameters and compared with the available experimental data. The optimal energy ranges for the production of $^{47}$Sc are determined by taking into account the thick-target yields and the purity of the product. The results show that both reactions can provide high yields and high purity of $^{47}$Sc. The feasibility of producing $^{47}$Sc with a hospital cyclotron is demonstrated by performing realistic simulations of the irradiation for both $^{50}$Ti(p,$\alpha$) and $^{49}$Ti(d,$\alpha$) reactions.}

\maketitle
\section{Introduction}
\label{intro}
Scandium-47 ($^{47}$Sc) is a radioisotope with potential applications in nuclear medicine, especially for targeted radiotherapy of cancer. It decays with a half-life of 3.349 days, emitting both $\beta^-$ particles and $\gamma$ rays, which are suitable for therapy and imaging, respectively. Moreover, it can form stable complexes with various biomolecules, such as peptides, antibodies, and nanoparticles, that can deliver it to specific tumor sites \cite{Domnanich17}.  However, the production of $^{47}$Sc is challenging, as it requires either neutron irradiation of $^{47}$Ti targets in nuclear reactors \cite{Soliman20} or cyclotron irradiation of protons on enriched targets (e.g., $^{48}$Ca targets, see \cite{Misiak17}). As of today none of the attempted production routes have proven feasibile in full. Therefore, finding a good production route for $^{47}$Sc is important for making it more available and accessible for nuclear medicine applications. 

In this paper, we first use a statistical description to visualize the trend 
and dispersion arising in the cross section when applying the variety of 
theoretical models included in mass-production nuclear reaction codes such as 
TALYS \cite{Koning23}. For an outline on this statistical pathway to represent the model-trend and model-dispersion or variability, see Sect. 6.1.2 of Ref.\cite{Ballan23}. Then, for an accurate evaluation of the production routes,  we propose to use genetic algorithms (GAs) to optimize the parameters of nuclear reaction models that describe the production of $^{47}$Sc from different reactions. GAs are a type of evolutionary computation that mimic the process of natural selection to find optimal solutions to complex problems \cite{Sean12}. They can be used to optimize the parameters of nuclear reaction models \cite{Wei19}, with the aim to describe production cross sections as accurately as possible \cite{Hilaire21}. Nuclear reaction models are important for understanding various phenomena in nuclear physics, such as nuclear structure, nuclear astrophysics, nuclear fusion, nuclear fission, and can have a very important impact on nuclear medicine applications. However, finding the best parameters for these models is often challenging and time-consuming, as they depend on many factors and uncertainties. GAs can help overcome these difficulties by exploring a large and diverse search space of possible parameter values and selecting and combining the most promising ones based on their fitness or performance. In this work we optimize the parameters of nuclear reaction models for the production of $^{47}$Sc from two different reactions: $^{50}$Ti(p,$\alpha$)$^{47}$Sc  and $^{49}$Ti(d,$\alpha$)$^{47}$Sc. We compare the results of our optimization with experimental data. 

\section {The nuclear reaction $^{50}$Ti(p,x)$^{47}$Sc}

\begin{figure}[h]
\centering
A~\includegraphics[width=0.4\textwidth]{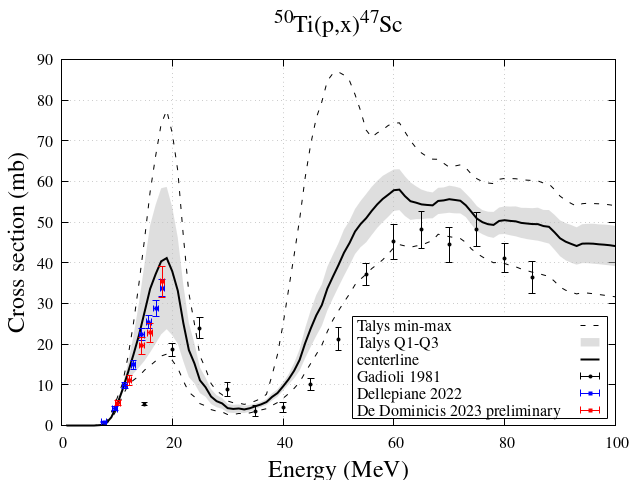}
\includegraphics[width=0.4\textwidth]{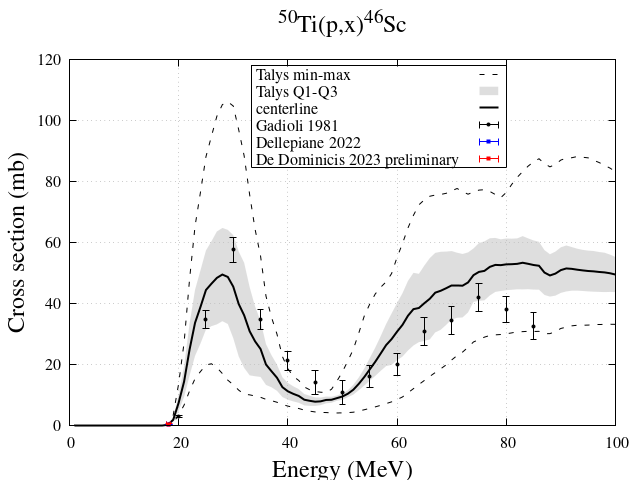}~B
C~\includegraphics[width=0.4\textwidth]{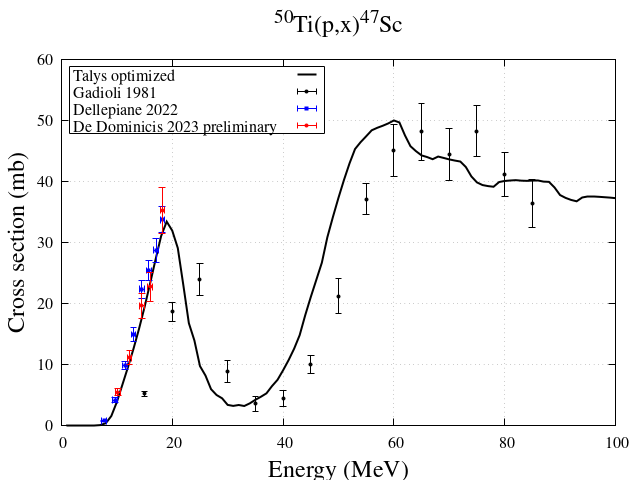}
\includegraphics[width=0.4\textwidth]{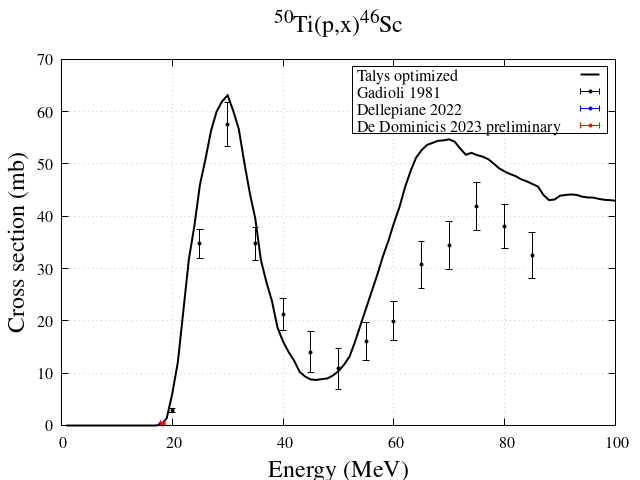}~D

\caption{Statistical description of $^{47}$Sc and $^{46}$Sc cross sections, in panels A and B, respectively. The dashed lines are the min-max values of the Talys models and the gray area denotes the interquartile band, measuring the model variability. The solid black line, i.e. the centerline of the band, represents the "trend" of the cross section, averaged over the models. In panels C and D, the corresponding optimized curves are compared with the measured cross sections.}
\label{p-ti50_xs47_ALL}
\end{figure}

The first reaction considered for $^{47}$Sc production involves the use of enriched $^{50}$Ti targets and proton beams. Fig. \ref{p-ti50_xs47_ALL} (panel A) describes the relevant cross sections calculated with TALYS and compares it to the available experimental data. The black dataset is the old measurements by Gadioli et al.\cite{Gadioli81}. Newest data by Dellepiane et al.\cite{Dellepiane22}  and by De Dominicis et al.\cite{DeDominicis23, DeDominicis23b, Pupillo23} are also reported with blue and red dots, respectively. The last two data sets are in contrast with the previous data and open new promising perspectives for this reaction. None of the TALYS models was able to reproduce the peak measured by Gadioli at around 25 MeV: the trend the TALYS models, expressed by the solid black line representing the medium between 1st and 3rd quartile of all the TALYS model calculations, indicates the peak occurring slightly before 20 MeV.  Instead, the two new data sets are mutually consistent and show good agreement with the calculated cross sections with TALYS. Likely, the 1981 Gadioli measurements at low-energy (lower than 30 MeV) have an issue. The analysis has been extended to the cross section of $^{46}$Sc production, one of the main contaminants. Fig.\ref{p-ti50_xs47_ALL} (panel B) shows both the theoretical cross section and the experimental data. For both \cite{Dellepiane22} and \cite{DeDominicis23} only one energy point has been measured in correspondence of the threshold energy of the reaction.

While panels A and B of Fig. \ref{p-ti50_xs47_ALL} provide an overall statistical picture of the trend and variation of TALYS model calculations, it is important to find a single, optimized calculation that describe accurately the data. Thus, it can be utilized to predict the irradiation parameters in nuclear medicine application. We have found that the most adequate combination of models corresponds to PE3 - LD4 with the following modification of the  $c$ and $p$ level density parameters for the compound nucleus $^{47}$Sc:  c = 0.0 MeV$^{-1/2}$ and p = 0.5 MeV. The corresponding results for $^{47}$Sc and $^{46}$Sc production are reported in panels C and D, respectively, with a solid line. The optimization of the level density of the compound $^{47}$Sc nucleus improves also the cross section for $^{46}$Sc production. 
Although in both cases the reproduction of the cross section at higher energies are not optimal, one should observe that at those energies only older data\cite{Gadioli81} are available and they might not be that precise to optimize the cross sections.
In addition, this work is particularly focused on low energies, where the 
$^{46}$Sc cross section remains negligible, preserving the purity of the $^{47}$Sc 
production. The energy range of interest is limited to below 18 MeV.

\begin{figure}[h!]
\begin{center}
A~\includegraphics[width=0.4\textwidth]{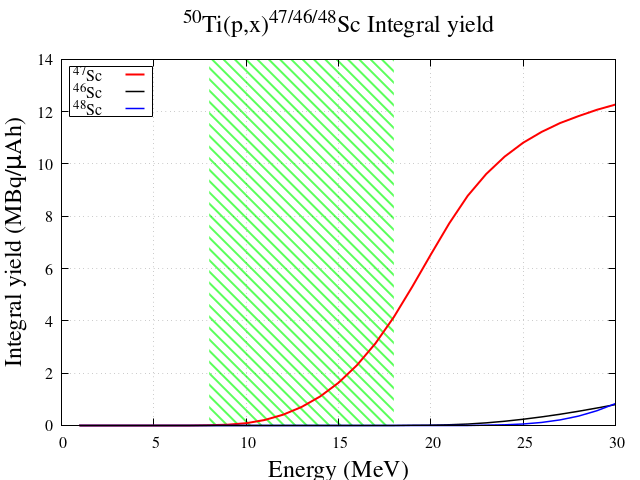}
\includegraphics[width=0.4\textwidth]{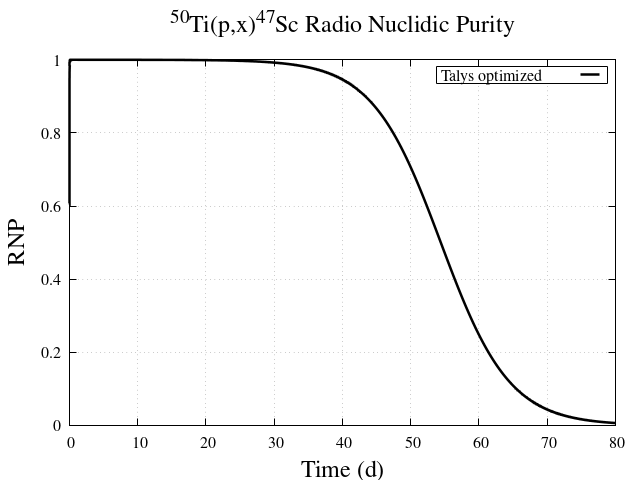}~B
\end{center}
\caption{Panel A denotes the integral yields of $^{47}$Sc, $^{46}$Sc, and $^{48}$Sc, and the shaded area highlights the production energy window. The time evolution of the radionuclidic purity is given in panel B. }
\label{p-ti50_yield_e_rnp}
\end{figure}

Based on the optimized cross section, the evaluation of yields was conducted 
for an irradiation duration of 1 hour and a current of 1 $\mu$A.
The comparison of $^{47}$Sc yields obtained with the optimized cross-section modeling is presented in panel A of Fig.\ref{p-ti50_yield_e_rnp}. The green area highlights the selected energy range, 8-18 MeV. The analysis of the yield has been performed also for the main contaminants, $^{46}$Sc and $^{48}$Sc. The red line refers to $^{47}$Sc yield, the black curve to $^{46}$Sc, and the blue one to $^{48}$Sc.
Clearly, the contribution from contaminants is suppressed within the highlighted window, and this affects positively the radionuclidic purity, which remains above 99\% for several weeks. In panel B, its time evolution is reported.


\section {The nuclear reaction $^{49}$Ti(d,x)$^{47}$Sc}

An alternative nuclear reaction that employs enriched $^{49}$Ti targets has been also investigated.
While we found that proton beams do not offer suitable energy ranges due to 
excessive contamination of both $^{46}$Sc and $^{48}$Sc, the use of deuteron beams on 
$^{49}$Ti is an interesting and promising channel to examine.
However, for this reaction, only one outdated experimental data set is 
available in the literature. This calls for new measurements to better 
understand the potential and efficacy of the production channel, providing an 
opportunity for a deeper investigation into the theoretical models used to 
describe this nuclear reaction.

To ensure the actual feasibility of the route it is necessary to perform a theoretical study considering the production yields of $^{47}$Sc and contaminants, and the level of purity that can be obtained.

\begin{figure}[h!]
\begin{center}
A~\includegraphics[width=0.4\textwidth]{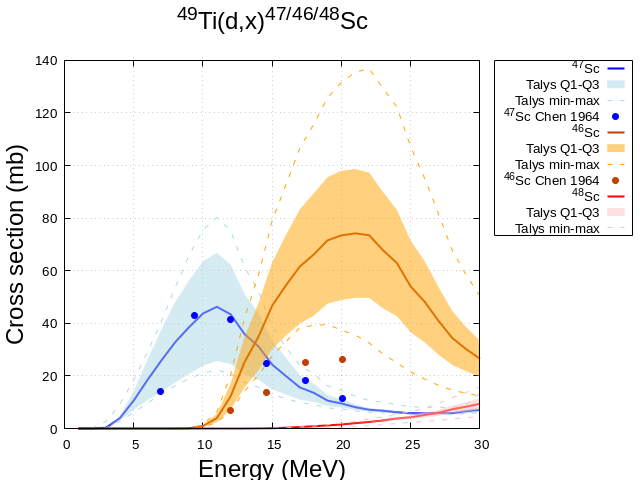}
\includegraphics[width=0.4\textwidth]{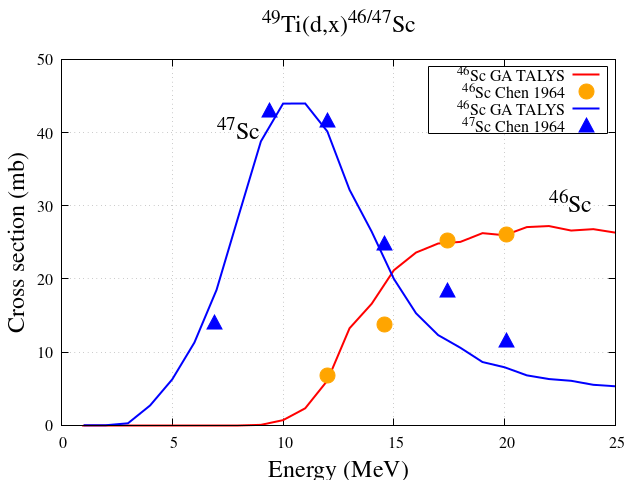}~B
C~\includegraphics[width=0.4\textwidth]{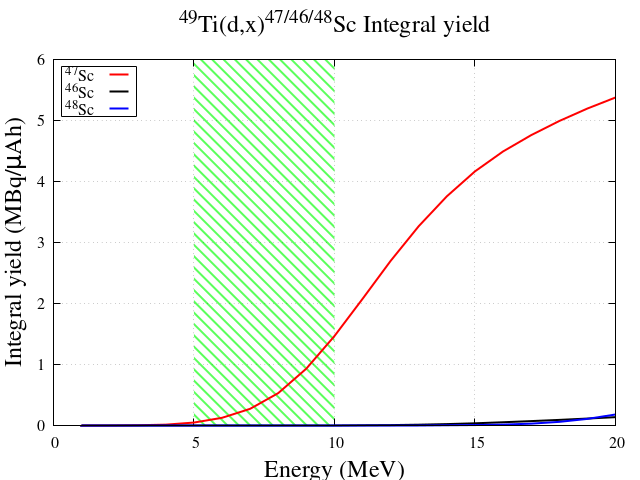}
\includegraphics[width=0.4\textwidth]{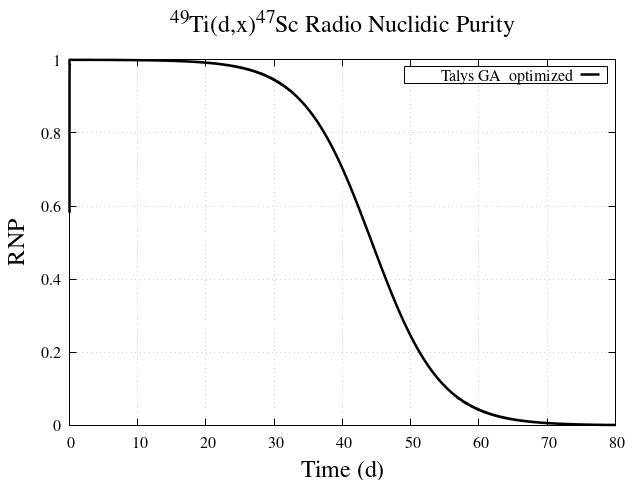}~D

\end{center}
\caption{Comparison of $^{47}$Sc (blue), $^{46}$Sc (orange), and $^{48}$Sc (red) cross sections using 
the statistical representation (panel A), in the case of deuteron beams on $^{49}$Ti target. The cross sections determined with GA optimization are shown in panel B. Integral yields determined from the 
GA-optimized cross sections are given in panel C. The time evolution of the corresponding radionuclidic purity is given in panel D.
}
\label{d-ti49_comparison-ALL}
\end{figure}

Fig. \ref{d-ti49_comparison-ALL} (panel A) illustrates the statistical cross section of the radionuclide $^{47}$Sc (in blue) and its main contaminants, $^{46}$Sc (orange) and $^{48}$Sc (red). The variability described by the TALYS models is wide, especially for the $^{46}$Sc cross section where the min-max range spans almost 100 mb. For these reactions the only available data are from Chen et al.\cite{Chen64}. An evident poor agreement exists between measured and theoretical cross section for the $^{46}$Sc case. However, focusing on the contaminant $^{46}$Sc, the reaction starts at around 10 MeV and it is well described there by a thin band. If one trusts the $^{46}$Sc result at least around 10 MeV it could be possible to limit the production of this contaminant. By restricting the energy range up to a maximum of 10 MeV, one will be in the condition where the peak of $^{47}$Sc occurs, ensuring maximum production of $^{47}$Sc minimal contamination of $^{46}$Sc. Thus, low-energy deuterons on enriched $^{49}$Ti targets is a promising route, with a significant $^{47}$Sc production potentially pure from contaminants. Hence, the possibility of using hospital cyclotrons providing 10 MeV deuterons is quite interesting. 

To improve the reproduction of the cross sections and achieve a good 
agreement with the measurements, enabling a more accurate prediction of yields 
and activities, it is essential to rely on the limited information gathered 
from the preliminary analysis. An optimization strategy by means of GA has been carried out. The TALYS combination of models PE3–LD6 has been identified as the most convenient to define the new optimized curves, since without any change in the free level-density parameters the cross sections were closest to the Chen data. Using GA, the optimized level density parameters were set to c = 1.573 MeV$^{-1/2}$ and p = 0.390 MeV for $^{46}$Sc compound nucleus, and c = -0.029 MeV$^{-1/2}$ and p = 1.327 MeV for the $^{47}$Sc one. The resulting cross sections are shown in panel B of 
Fig.\ref{d-ti49_comparison-ALL}, and they closely reproduce the experimental data.

Finally, the evaluation of the radionuclidic purity (panel D in Fig.\ref{d-ti49_comparison-ALL}) indicates a value higher than 99\% for almost 20 days, an outcome quite adequate for nuclear medicine applications.  In addition, the isotopic purity has been computed, revealing that it reaches 
and maintains a high value within 10 days. This suggests that this production 
route is essentially carrier-free, as it shows no significant presence of 
contaminants, including the long-lived or stable ones.



Since the 5 - 10 MeV window is below the rise in $^{46}$Sc cross section and in correspondence of $^{47}$Sc peak, it has been selected as an optimal energy-range for target irradiation.
Panel C of that figure shows the calculated integral yields obtained from the optimized cross section considering the standard irradiation conditions of T$_{irr}$ = 1h and I = 1 $\mu$A, and the green band delimits the selected energy to optimize the production. The curves in that panel denote the yields of $^{47}$Sc and of the two main contaminants. It is evident that the contamination by $^{46}$Sc and $^{48}$Sc can be considered negligible in the selected range. 

To conclude this Section, the nuclear reaction $^{49}$Ti(d,x) is a promising route thanks to the possibility to produce the radionuclide $^{47}$Sc almost pure with a very minimum contamination by $^{46}$Sc. Indeed, the theoretical results indicate a radionuclidic purity above the 
Pharmacopoeia's recommended threshold of 99\%. In order to improve the initial agreement of the measured and calculated cross sections, further analysis should be carried out. To this purpose the need of new experimental data is clear and this would validate the models and would allow to refine their tuning parameters.

\subsection{Limitations and perspectives}

As mentioned earlier, this reaction has been studied experimentally by Chen and Miller in 1964 \cite{Chen64}, who measured the cross sections for different deuteron energies. However, their data have some limitations that need to be addressed by future research.

Their target was not fully enriched in $^{49}$Ti, but contained a significant amount of $^{48}$Ti as well. This means that the cross sections they obtained are not pure for the $^{49}$Ti(d,$\alpha$)$^{47}$Sc reaction, but include contributions from other reactions involving $^{48}$Ti, such as $^{48}$Ti(d,$\alpha$)$^{46}$Sc. These reactions may affect the accuracy and precision of the cross-section measurements. Therefore, it would be desirable to repeat the experiment with a target that has a higher enrichment of $^{49}$Ti, or to correct the data for the presence of $^{48}$Ti.

Another limitation is that the nuclear reaction models used to describe this reaction do not properly account for the deuteron break-up contribution, which is an important process in deuteron-induced reactions. The deuteron break-up contribution refers to the possibility that the deuteron splits into a proton and a neutron before or after interacting with the target nucleus, resulting in different contributions to final states, compared to the direct reaction. However, most of the nuclear reaction codes used to calculate the cross sections for this reaction do not include this contribution by default. Therefore, further work is needed to improve the nuclear reaction models and include the deuteron break-up contribution in a more realistic way, as suggested by Avrigeanu \cite{Avrigenau23}.

To sum, the production route: $^{49}$Ti(d,$\alpha$)$^{47}$Sc to produce $^{47}$Sc is a feasible and attractive option for obtaining this radioisotope, but it requires more experimental and theoretical investigation to overcome some of the limitations of the existing data and models. A new experiment with a highly enriched $^{49}$Ti target and a more accurate measurement of the cross sections would be valuable, as well as a better understanding and modeling of the deuteron break-up contribution in this reaction.

\section{Results and conclusive comments}

In Table \ref{tab:yield_d-ti49} the integral yield values are reported for the two production routes, comparing $^{47}$Sc and the two contaminants. Considering that $^{47}$Sc has an half-life of 3.349 $d$, the yield could be enhanced by increasing significantly the irradiation time before arriving at saturation, and/or the current. 

Both production routes are suitable for production of this radionuclide with purity suitable for medical applications. The production with deuterons require a 10-MeV beam, a clear advantage compared with the required 18-MeV proton beam for the $^{50}$Ti(p,$\alpha$) reaction. However, both productions can be performed with hospital-type cyclotrons. The reaction with protons, on the other hand, performs better in yield (almost three-times) and purity.



\begin{table}[!htb]
\begin{center}
\footnotesize

\begin{tabular}{|c|c|}

\hline
$^{50}$Ti(p,$\alpha$) [Yield MBq/($\mu$A $\cdot$ h) ]
&
$^{49}$Ti(d,$\alpha$) [Yield MBq/($\mu$A $\cdot$ h) ]
\\
\hline
\begin{tabular}{|c|c|c|}
\hline
\multicolumn{3}{|c|}{E$_{max}$-E$_{min}$  = 18 -- 8 MeV}\\
\hline
$^{47}$Sc & $^{46}$Sc & $^{48}$Sc \\
\hline
4.138 & 8.35E-05 & 0.0 \\
\hline
\end{tabular}

&

\begin{tabular}{|c|c|c|}
\hline
\multicolumn{3}{|c|}{E$_{max}$-E$_{min}$  = 10 -- 5 MeV}\\
\hline
$^{47}$Sc & $^{46}$Sc & $^{48}$Sc \\
\hline
1.418 & 2.14E-04 & 0.0 \\
\hline
\end{tabular}

\\
\hline
\end{tabular}

\caption{Integral yield of $^{47}$Sc and contaminants calculated for T$_{irr}$ = 1h, I = 1 $\mu$A, for both production routes considered in this work.}
\label{tab:yield_d-ti49}
\end{center}
\end{table}






\begin{thebibliography}{}
%
%
\bibitem{Domnanich17} Domnanich K. A., et al.,  EJNMMI Radiopharm. Chem. \textbf{2}, 5 (2017)
\bibitem{Soliman20} Soliman M. A., et al.,   J Radioanal Nucl Chem \textbf{324}, 1099 (2020)
\bibitem{Misiak17} Misiak, et al., J Radioanal Nucl Chem \textbf{313}, 429 (2017)
\bibitem{Koning23} Koning A., et al., Eur. Phys. J. \textbf{A59} 131 (2023)
\bibitem{Ballan23} Ballan M., et al., Eur. Phys. J. Plus \textbf{138} 709 (2023) (Sect. 6.1.2)
\bibitem{Sean12} Sean L., \textit{Essential in Metaheuristics (Second Edition)} (lula.com, 2012) 242
\bibitem{Wei19} Wei Z., et al., in IEEE Access, \textbf{7}, 66084 (2019)
\bibitem{Hilaire21} Hilaire S. and Goriely S., Springer Proceedings in Physics \textbf{254} 3 (2021)
\bibitem{Gadioli81} Gadioli E. et al.
Z. Phys. A-Hadron. Nucl.,
\textbf{301}, 289–300 (1981).

\bibitem{Dellepiane22}
Dellepiane G., et al.
Applied Radiation and Isotopes,
\textbf{189}, 110428
(2022)

\bibitem{DeDominicis23}
De Dominicis L., Private Communication (2023)

\bibitem{DeDominicis23b}
De Dominicis L., et al.  J. Phys.: Conf. Ser. \textbf{2586} 012128 (2023)

 \bibitem{Pupillo23}
Pupillo G., et al., J. Phys.: Conf. Ser. \textbf{2586} 012118 (2023)

\bibitem{Chen64} Chen K. L.  and  Miller J. M.,
Phys. Rev. \textbf{134}, B1269 (1964)

\bibitem{Avrigenau23}
Avrigenau M., \textit{Due consideration of breakup and stripping mechanisms within (d; p), (d; 2p), and (d; xn) reactions} presented at this Conference (2023).
\end{thebibliography}
%
%

\end{document}